\documentclass{appolb}
\usepackage{epsfig}
\begin{document}
\title{Evidence for an Anti-charmed Baryon State
\thanks{Presented at the Cracow Ephipany Conference on Hadron Physics,
Cracow, Poland, January 6-8, 2005}%
}
\author{Chr. RISLER
\address{for the H1 collaboration}
\address{DESY, Notkestrasse 85,D-22607 Hamburg, Germany, \\
        email: risler@mail.desy.de }
%\and
%the Name(s) of other Author(s)
%\address{and their affiliation}
}
\maketitle
\begin{abstract}
We report on the observation of a narrow resonance 
in $D^{*-}p$ and $D^{*+} \bar p$ invariant mass
combinations in deep-inelastic $ep$ scattering at centre-of-mass energies 
of 300 and 320 GeV at HERA. 
The mass of the resonance is measured to be 
$3099 \pm 3 ({\rm stat.}) \pm 5 ({\rm syst.}) \, {\rm MeV}$, 
the Gaussian width of $12 \pm 3 ({\rm stat.}) \, {\rm MeV}$ 
is compatible with the experimental resolution. 
The state can be interpreted as an anti-charmed baryon with minimal 
constituent quark composition $uudd\bar c$, together with the charge conjugate.
\end{abstract}
\PACS{14.20.Lq,14.80.-j}
  
\section{Introduction}
In the last 2 years several experiments \cite{thetaKn} have reported evidence 
of a narrow baryonic resonance with strangeness $S=+1$ 
in the invariant mass of $K^+n$ combinations.
These resonances can  be interpreted as candidates for a strange pentaquark $\theta^+$,
the minimal constituent quark content being $uudd\bar s$.
These measurements were supported by similar observations \cite{thetaKp} 
in the $K^0_S p(\bar p)$ spectrum, although this channel does not 
allow for the observation of exotic quantum numbers, since the $K^0_S$ is a 
linear combination of strangeness $S=+1$ and $S=-1$ states. However, there are also
a number of high-energy experiments 
that do not confirm the observation of $\theta^+$ candidates \cite{nonobsthpl}.
Also evidence for the pentaquark cascade states, $\Xi^{--}_{5}$ and $\Xi^{0}_5$ 
with strangeness $S=-2$, has been reported \cite{NA49}.
The observations of possible $\theta^+$ candidates 
have motivated H1 to search for a  charmed pentaquark. The possible existence
of such states had been discussed before \cite{cpqtheory}.
The clearest charm signal is seen in the decay of the $D^{*\pm}$. 
Therefore a search for resonances in the $D^{*}p$ invariant mass
spectrum was performed and evidence for a narrow baryonic resonance 
in the $D^{*-}p$ spectrum and its charge conjugate was found.
In the following this analysis, which is published by the H1 collaboration \cite{H1cPQ}, 
is briefly described.\\

\section{Event Selection}
The analysed data were collected with the H1 detector in the
years 1996 to 2000 and correspond to an integrated 
luminosity of $75 \, {\rm pb}^{-1}$. A detailed description of the H1 
detector can be found elsewhere \cite{H1det}.
Deep-inelastic scattering (DIS) events were selected by 
requiring a reconstructed scattered electron in the backward calorimeter
of H1 and an exchanged photon virtuality of $Q^2>1 \, {\rm GeV}^2$.
The kinematic range is further restricted to values of the inelasticity $y$  
of $0.05<y<0.7$ in order to ensure substantial hadronic final state energies
in the central detector region.
As an independent sample photoproduction ($\gamma p$) events are used, were the 
scattered electron emits a quasireal photon and is not detected in 
the central detector but escapes in the beam pipe. Photoproduction events
are selected by requiring $Q^2 < 1 \, {\rm GeV}^2$.
\subsection{$D^*$ and proton reconstruction}
\begin{figure}[btp]
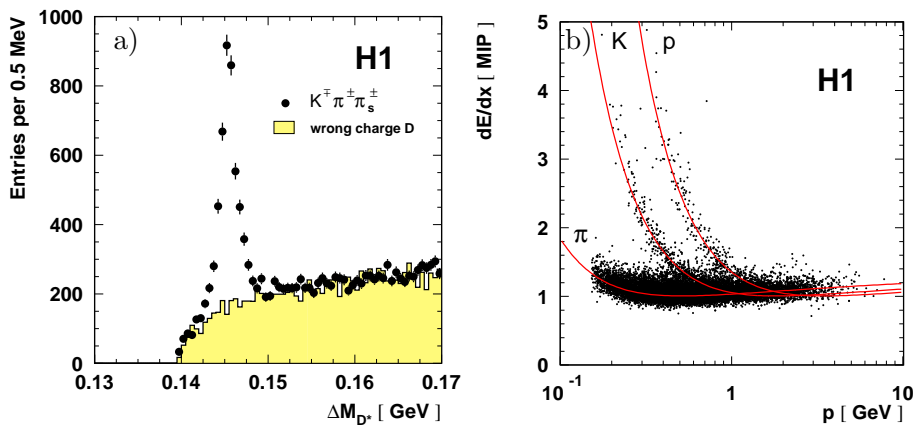

  \setlength{\unitlength}{1cm}
  \begin{center}
    \begin{picture}(12,6)
      \put(1.5,5){a)}
      \put(0,0){
        \includegraphics[width=6.0cm]{fig1a.epsi}}
      \put(7.5,5){b)}
      \put(6.2,0){
        \includegraphics[width=5.8cm]{fig1b.epsi}}
    \end{picture}
  \end{center}
  \caption{\small\label{Dstsignal}
    (a) Distribution of the mass difference $\Delta M(D^*)= M(K\pi\pi)-M(K\pi)$ for DIS events ($Q^2 > 1 \, {\rm GeV^2})$. The background
    under the $D^*$ signal is described by the wrong charge D combinations (see text). 
    (b) Specific energy loss due to ionization ${\rm d}E/{\rm d}x$ versus momentum for proton candidate tracks. }
\end{figure}

The decays of the charmed $D^*$ mesons are reconstructed via the decay channel,
$$D^{*\pm} \rightarrow D^0 \pi^{\pm}_{sl} 
\rightarrow (K^{\mp} \pi^{\pm}) \pi^{\pm}_{sl}\quad ,$$
which provides a particularly clean $D^*$ signal, although the branching ratios are low.
A mass difference technique is applied, $\Delta M(D^*) = M(K\pi\pi)-M(K\pi)$,
in order to improve the mass resolution.
Candidates of three particle compbinations are used if the reconstructed
$D^0$ mass $M(K\pi)$ 
is close to its nominal value \cite{pdg}, 
        $$ | M(K\pi)-M(D^0)_{PDG}|<60 \, {\rm MeV} \quad .$$
After applying further cuts on the transverse momenta,
$p_T(D^*)>1.5 \, {\rm GeV}$ and 
$p_T(K)+p_T(\pi) > 2 \, {\rm GeV}$, 
the pseudorapidity $-1.5 < \eta(D^*) < 1.0$
and the production elasticity 
$z(D^*)=(E-p_z)_{D^*}/2yE_e>0.2$ of the $D^*$ a good signal to 
background ratio in the $\Delta M(D^*)$ distribution is achieved, 
as shown in  fig.~\ref{Dstsignal}a, yielding about 3500 $D^*$ candidates.
The background is mainly due to combinatorics, not involving any charm decays 
and can be estimated from the data by using the wrong charge D
combinations, where instead of the oppositely charged $K$ and $\pi$ candidate tracks 
forming the $D^0$, two tracks of the same charge are selected forming a 
doubly charged pseudo $D$ which further is combined with a slow pion track.
The background under the $D^*$ signal is
well described by such wrong charge D combinations.\\

The $D^*$ candidates are combined further with  charged tracks originating from
the primary vertex assigned the proton mass.
These proton tracks are selected using the measurement of the ionisation loss
${\rm d}E/{\rm d}x$ in the central drift chambers of H1. 
The average ${\rm d}E/{\rm d}x$ resolution for
minimal ionising particles is about 8\% \cite{steinhart}. 
An example of the dE/dx distribution
as a function of particle momenta is shown in fig.~\ref{Dstsignal}b. 
Bands for $\pi$, $K$ and protons are clearly visible.
The measurement is compared to the Bethe-Bloch-like parameterisation shown by
the solid lines. From the difference of the measurement and the parameterisation 
the likelihood probabilities for different particle hypotheses are calculated
which are used for particle identification.\\
\section{$D^* p$ signal}
\begin{figure}[btp]
  \setlength{\unitlength}{1cm}
  \begin{center}
    \begin{picture}(9,6)
      \put(0,0){
        \includegraphics[height=6cm]{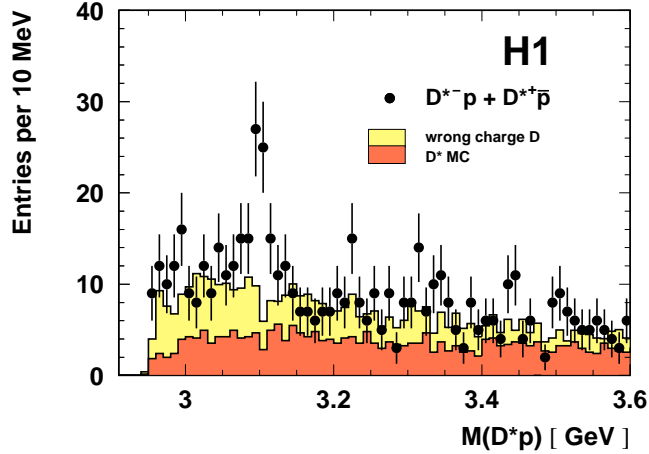}}
    \end{picture}
  \end{center}
  \caption{\small\label{Dstprsignaloppsign}
    Distribution in $M(D^*p)$ for opposite-charge $D^*p$ combinations. The data are compared with
    the sum of a non-charm contribution estimated using the wrong charge D combinations
    and a simulated charm contribution (see text).
    }
\end{figure}
\begin{figure}[btp]
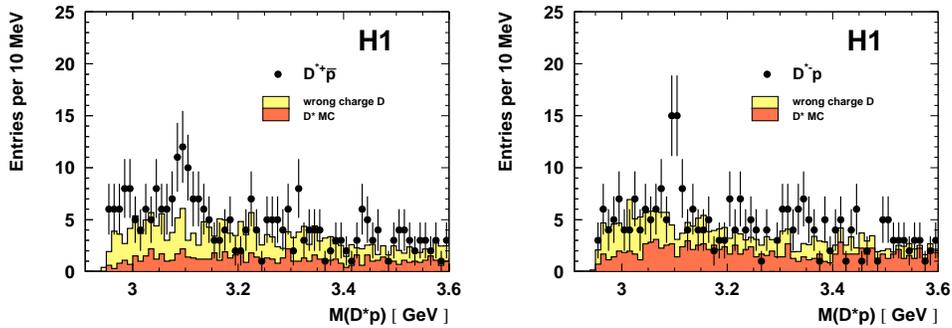

  \setlength{\unitlength}{1cm}
  \begin{center}
    \begin{picture}(12.5,5)
      \put(0,0){
        \includegraphics[width=6cm]{fig3a.epsi}}
      \put(6.5,0){
        \includegraphics[width=6cm]{fig3b.epsi}}
    \end{picture}
  \end{center}
  \caption{\small\label{Dstprsignalb}
    Distribution in $M(D^*p)$ for opposite-charge $D^*p$ combinations, 
    separately for $D^{*-}p$ (left) and 
    $D^{*+}\bar p$ (right). Background model as in fig.~\ref{Dstprsignaloppsign}.
  }
\end{figure}
\begin{figure}[btp]
  \setlength{\unitlength}{1cm}
  \begin{center}
    \begin{picture}(9,6)
      \put(0,0){
        \includegraphics[height=6cm]{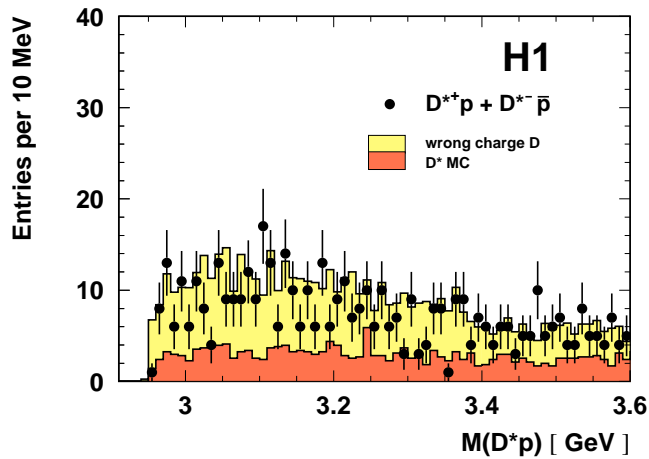}}
    \end{picture}
  \end{center}
  \caption{\small\label{Dstprsignallikesign}
    Distribution in $M(D^*p)$ for same-charge $D^*p$ combinations. Background model as in fig.~\ref{Dstprsignaloppsign}. }
\end{figure}

The selected $D^*$, fulfilling 
$$| \Delta M(D^*) - (M(D^0)_{PDG}-M(D^*)_{PDG})| < 2.5 \, {\rm MeV} \quad ,$$
are further combined with the proton candidates.
The invariant $D^* p$ mass is formed again exploiting 
the mass difference method, 
$M(D^*p)= M(K\pi\pi p)-M(K\pi\pi) + M(D^*)_{PDG}$.
The mass difference for the opposite-charge combinations  
$D^{*-}p$ and $D^{*+}\bar p$ is shown in fig.\ref{Dstprsignaloppsign}.
A narrow peak at $M(D^*p)\approx 3.1 \, {\rm GeV}$ is cleary visible.
This signal is observed with similar strength and compatible width
in the $D^{*-}p$ and $D^{*+}\bar p$ combinations separately (fig.\ref{Dstprsignalb}).
The background under the $D^*p$ signal can be reasonably described by the sum of two components:
a non-charm background estimated from the data using the above 
mentioned wrong charge D combinations and a charm contribution, where a 
real $D^*$ is combined with random proton tracks. 
The latter is estimated from Monte Carlo simulations of
the $D^{*}$ production in DIS events using the RAPGAP generator \cite{RAPGAP}.
No significant peak is observed 
in the invariant mass of the same-charge combinations, 
$D^{*+}p$ and $D^{*-}\bar p$,
shown in fig.\ref{Dstprsignallikesign}.
The data are compatible with the sum of the charm and the non-charm 
background.\\
\section{Signal Tests and Significance}
Extensive tests have been performed to examine the observed signal.
%%%%%%%%%%%%%%%%%%%%%%%%%%%%%%%%%%%%%%%%%%%%%%%%%%%%%%%%%%%%%%%%%%%%%%%%%%%%%%%%%%%%%%%%%%%
\begin{figure}[btp]
  \setlength{\unitlength}{1cm}
  \begin{center}
    \begin{picture}(6,6)
      \put(0,0){
        \includegraphics[width=6cm]{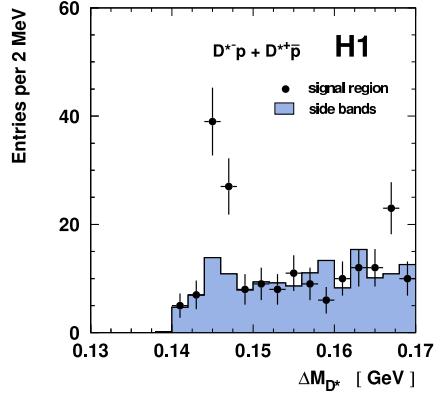}}
    \end{picture}
  \end{center}
  \caption{\small\label{backward}
    $\Delta M(D^*)$ distribution for events in a 30 MeV window around the signal in the opposite-charge
    $M(D^*p)$ distribution, with no requirement on $\Delta M(D^*)$, compared with the corresonding
    distribution from the $D^*p$ sideband regions, normalised according to the width of the
    mass window.
    }
\end{figure}
%Sideband studies confirmed that the $M(D^*p)$ signal region yields a richer $D^*$ content
%than the sideband regions, showing that the signal is due to $D^*$ mesons and not produced
%in the sidebands of the $\Delta M(D^*)$ distribution.\\
The $D^*$ content of the signal has been investigated by comparing the $D^*$ signals in the
signal and sideband regions of the $M(D^*p)$ distribution, using the full proton selection
but no requirement on $\Delta M(D^*)$. The $\Delta M(D^*)$ distribution is shown in fig.~\ref{backward}
for events in a $\pm 15$ MeV mass window around the $D^*p$ signal, $ 3085 < M(D^*p) < 3115$ MeV, compared
with the similar distribution from the sidebands. The $\Delta M(D^*)$ distribution from the sidebands
is scaled by a factor accounting for the different widths of the sideband and signal mass windows.
In the $\Delta M(D^*)$ region above the $D^*$ peak the distributions agree with each other in 
shape and normalisation. However, there is a clear difference around the $D^*$ peak position,
where the distribution from the signal region in $M(D^*p)$ overshoots that from the sidebands.
The signal region in $M(D^*p)$ is thus significantly richer in $D^*$ mesons than the sideband regions.\\

\begin{figure}[btp]
  \setlength{\unitlength}{1cm}
  \begin{center}
    \begin{picture}(6,6)
      \put(0,0){
        \includegraphics[height=6cm]{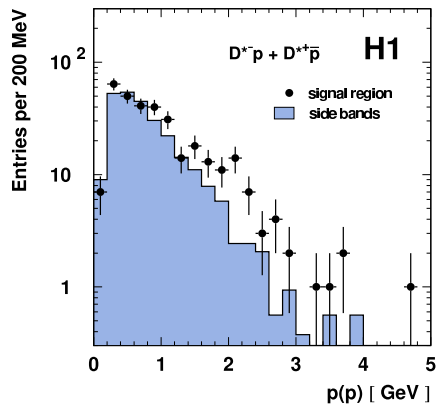}}
     % \put(6,0){
      %  \includegraphics[width=6cm]{fig6b.epsi}}
    \end{picture}
  \end{center}
  \caption{\small\label{protonmomenta}
    Momentum distribution of all proton candidates yielding $M(D^*p)$ values falling
    in the signal and sideband regions of the signal in $M(D^*p)$.
    %(b) $M(D^*p)$ invariant mass distribution
    %for a high momentum selection ($p(p) > 2 \, {\rm GeV}$) with no proton ${\rm d}E/{\rm d}x$ requirement. The data are compared
    %with the same background model as in fig.~\ref{Dstprsignaloppsign}.
    }
\end{figure}

%%%%%%%%%%%%%%%%%%%%%%%%%%%%%%%%%%%%%%%%%%%%%%%%%%%%%%%%%%%%%%%%%%%%%%%%%%%%%%%%%%%%%%%%%%%%%%%
The proton content of the signal has been tested in the following way. At low proton momenta, $p(p)< 1.2 \, {\rm GeV}$ 
a more stringent particle identification in ${\rm d}E/{\rm d}x$  can be applied.
In this proton enriched sample a clear peak at $M(D^*p)\approx 3100 \, {\rm MeV}$ is visible.
A harder momentum spectrum of the proton candidates in the signal region
of the $M(D^*p)$ distribution compared with its sidebands was observed, as shown in fig.\ref{protonmomenta}.
\begin{figure}[btp]
  \setlength{\unitlength}{1cm}
  \begin{center}
    \begin{picture}(9,6)
%      \put(0,0){
%        \includegraphics[width=6cm]{fig6a.epsi}}
      \put(0,0){
        \includegraphics[height=6cm]{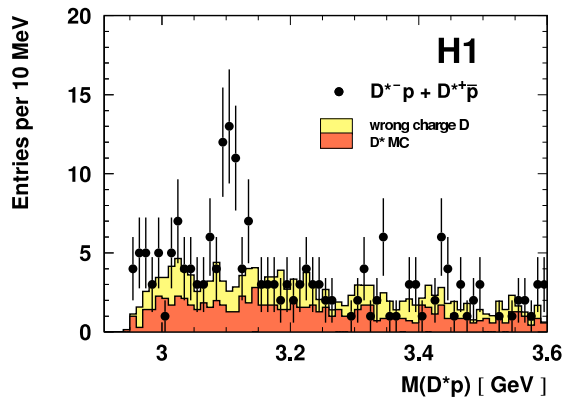}}
    \end{picture}
  \end{center}
  \caption{\small\label{protonmomentab}
    %(a) Momentum distribution of all proton candidates yielding $M(D^*p)$ values falling
    %in the signal and sideband regions of the signal in $M(D^*p)$. (b) 
    $M(D^*p)$ invariant mass distribution
    for a high momentum selection $p(p) > 2 \, {\rm GeV}$ with no proton ${\rm d}E/{\rm d}x$ reqiurement. The data are compared
    with the same background model as in fig.~\ref{Dstprsignaloppsign}.
    }
\end{figure}
This behaviour is expected for a two-body resonance decay, since for single charged particles
a steeply falling momentum distribution is expected, which is conserved when forming the combinatorial
background. Whereas in case of a real resonance the decay particles can be emitted in the direction of flight
of the original particle and therefore may have larger momenta in the laboratory frame.
Fig.\ref{protonmomenta} suggests that the signal to background ratio improves as the proton momentum increases.
In fig.~\ref{protonmomentab} the $M(D^*p)$ distribution is shown for momenta $p(p)> 2 \, {\rm GeV}$ 
without any particle identification requirement. A strong signal over a reduced background, well described
by the charm and non-charm background models, is observed. The peak position and width are compatible with those observed
with the standard selection.\\

%%%%%%%%%%%%%%%%%%%%%%%%%%%%%%%%%%%%%%%%%%%%%%%%%%%%%%%%%%%%%%%%%%%%%%%%%%%%%%%%%%%%%%%%%%%%%%%%%%%%%%%%%%%%%%%%%
Possible reflections from other resonances have been studied by investigating mass distributions and correlations
under different mass hypothesis for the $K, \pi$ and proton candidate tracks. None of these studies gave an explanation
for the observed structure. In particular the orbitally excited $D_1$ and $D_2$ states decaying to $D^*\pi$ do not give
a significant contribution to the peak.
An independent photoproduction sample shows a $D^*p$ resonance structure which is compatible with the one
seen in DIS events.
Furthermore all events have been visually scanned without discovering
any anomalies in the reconstruction of the events neither in the signal nor in the background regions.\\

\begin{figure}[btp]
  \setlength{\unitlength}{1cm}
  \begin{center}
    \begin{picture}(9,6)
      \put(0,0){
        \includegraphics[height=6cm]{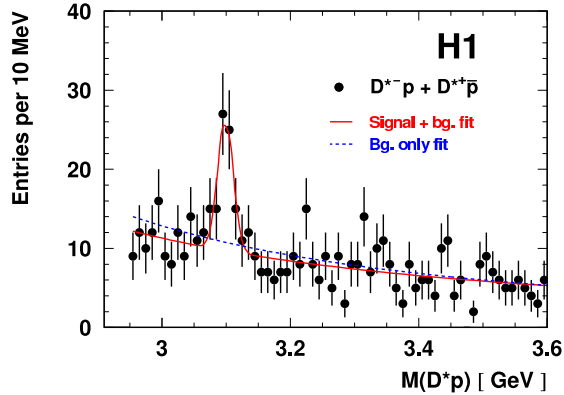}}
    \end{picture}
  \end{center}
  \caption{\small\label{Dstprsignalfit}
    $M(D^*p)$ distribution from opposite-charge $D^*p$ combinations compared with the result of a fit
    including both signal and background distributions (solid line) and with a fit including only the
    background component (dashed line).
    }
\end{figure}

The signal in fig.~\ref{Dstprsignaloppsign} is quantified by fitting the mass distribution 
by a gaussian together with a background function, $\alpha[M(D^*p)-M(D^*)]^\beta]$, 
as shown in fig.~\ref{Dstprsignalfit}.
The resulting peak position is $3099 \pm 3 {\rm (stat) \,  MeV}$ with a systematic uncertainty 
of 5 MeV. The gaussian width
is $12\pm 3 {\rm (stat) \, MeV}$, 
which is compatible with the experimental resolution of $7\pm2$ MeV.
The fit yields $N_S=50.6 \pm 11.2$ signal events under the peak, which corresponds to
a raw ratio of $N_S/N(D^*)$ of $1.46\pm0.32\%$ of the $D^*$ yield, not corrected for
detector acceptances. In a 2$\sigma$ window around the peak position $N_B=45.0 \pm 2.8$ 
background events are observed.
In order to estimate the significance in a more conservative approach the full distribution
is fitted by the background shape without adding a gaussian to describe the signal (dashed line
in fig.~\ref{Dstprsignalfit}), yielding  $N_B=51.7 \pm 2.7$ background events in the above mentioned
2$\sigma$ mass window. The poisson probability for an expectation 
of $N_B=51.7$ events to fluctuate to the observed 95 events or more is $ 4 \times 10^{-8}$, which corresponds
to a significance of $5.4\sigma$ in terms of gaussian standard deviations.
\section{Discussion}
A clear narrow resonance is observed in the invariant mass of $D^*p$ combinations in deep-inelastic scattering events
for $Q^2>1 \, {\rm GeV}^2$ with a width 
compatible with the detector resolution. The signal can not be explained in terms of
reflections or reconstruction failure and was robust under several tests. 
The background model well describes the observed background in the $D^*p$ mass distributions.
The observation of the narrow structure seen in DIS events is also observed in the analysis of an indendent photoproduction
sample.
However, several experiments searched for this structure and could not confirm its observation.
Negative results where reported by e.g. the FOCUS, ALEPH, CDF and BELLE collaboration \cite{nonobs} 
in different reactions and different phase space regions as at H1, which complicates the direct comparison.
The ZEUS collaboration \cite{nonobs2} found no signal in the same reaction and similar kinematic cuts and phase space.
The experimental discrepancy needs further investigation and clarification, possibly from the 
upcoming HERA-II data.
\section{Conclusion}
A narrow resonance is observed in $D^{*-}p$ and $D^{*+}p$ combinations at
$M(D^*p)=3099 \pm 3 {\rm (stat)} \pm 5 {\rm (syst) \, MeV}$ with a gaussian width
of $12\pm 3 {\rm (stat) \, MeV}$, compatible with the experimental resolution. The statistical significance
is estimated to be 5.4$\sigma$. The observed baryonic resonance may be interpreted as anti-charmed
baryon with minimal constitutent quark content $uudd\bar c$.

\end{document}